\documentstyle[11pt,paspconf,epsf]{article}

\begin{document}
\raisebox{0cm}[0cm][0cm]{\makebox[0cm][l]
{\hspace{-3.5cm}\parbox{16cm}{\em
To appear in "The Low Surface Brightness Universe", IAU Coll 171, eds. J.I.
Davies et al., A.S.P. Conference Series
}}}

\title{The Fornax
Spectroscopic Survey --- Low Surface Brightness Galaxies in Fornax}

\author{M.J. Drinkwater}
\affil{Physics, University of New South Wales, Sydney 2052, 
Australia}
\author{S. Phillipps, J.B. Jones}
\affil{Physics, University of Bristol, Tyndall Avenue, Bristol BS8 1TL, UK}

\begin{abstract}
{\em The Fornax Spectroscopic Survey} is a large optical spectroscopic
survey of {\em all} 14~000 objects with $16.5<B_J<19.7$ in a 12
deg$^2$ area of sky centered on the Fornax Cluster. We are using the
400-fibre Two Degree Field spectrograph on the Anglo-Australian
Telescope: the multiplex advantage of this system allows us to observe
objects conventionally classified as ``stars'' as well as
``galaxies''. This is the only way to minimise selection effects
caused by image classification or assessing cluster membership.

In this paper we present the first measurements of low surface
brightness (LSB) galaxies we have detected both in the Fornax Cluster
and among the background field galaxies. The new cluster members include
some very low luminosity ($M_B\approx-11.5$\,mag) dwarf ellipticals,
whereas the background LSB galaxies are luminous
($-19.6<M_B<-17.0$\,mag) disk-like galaxies.
\end{abstract}


\section{Introduction: The Fornax Spectroscopic Survey}

Several remarks have already been made at this Colloquium about the
difficulty of performing optical redshift surveys of low surface
brightness (LSB) galaxies, in particular using fibre-fed spectroscopy,
despite the pressing need for redshifts for these objects. It is hard
to obtain optical spectra of LSB galaxies at the best of times. The
limited apertures of fibre spectrographs and problems with sky
subtraction would normally be thought to make matters even
worse. However by using a system with a very large number of fibres
like the 400-fibre Two Degree Field (2dF) on the Anglo-Australian
Telescope (AAT) these limitations are outweighed by the multiplex advantage
of the system.

{\em The Fornax Spectroscopic Survey} (FSS) is designed to sample the
largest possible range in surface brightness, including high surface
brightness (HSB) as well as LSB galaxies. It does this by targeting
{\em all} objects in a region of sky centred on the Fornax Cluster. No
morphological information is used in the target selection, so objects
conventionally classified as ``stars'' are included as well as
``galaxies''. Such a complete survey of all objects is the only way to
minimise selection effects caused by image classification or assessing
cluster membership. Previous attempts at ``all-object'' surveys have
been limited to small areas.  Morton, Krug \& Tritton (1985)
obtained spectra of all 606 star-like objects brighter than $B=20$ in
an area of 0.31 deg$^2$ and Colless et al.\ (1991) extended their
galaxy survey by measuring spectra of 117 compact objects with
$21<B_J<23.5$ in a 0.1 deg$^2$ area.  {\it The Fornax Spectroscopic
Survey} will cover four 2dF fields (totaling 12 deg$^2$) centered on
the Fornax Cluster. It will include all 14,000 objects in the
magnitude range of $16.5<B_J<19.7$ (and somewhat deeper for unresolved
images).

We selected our targets from an APM (Irwin et al.\ 1994) scan of the
blue and red sky survey plates which provided accurate positions,
image classifications and photographic $B_J$ and $R$ magnitudes
(optimised for stellar profiles). The galaxy $B_J$ magnitudes were
taken from Davies et al.\ (1988), as were estimates of the surface
brightness and image scale length.

\begin{figure}[!htb]
\plotfiddle{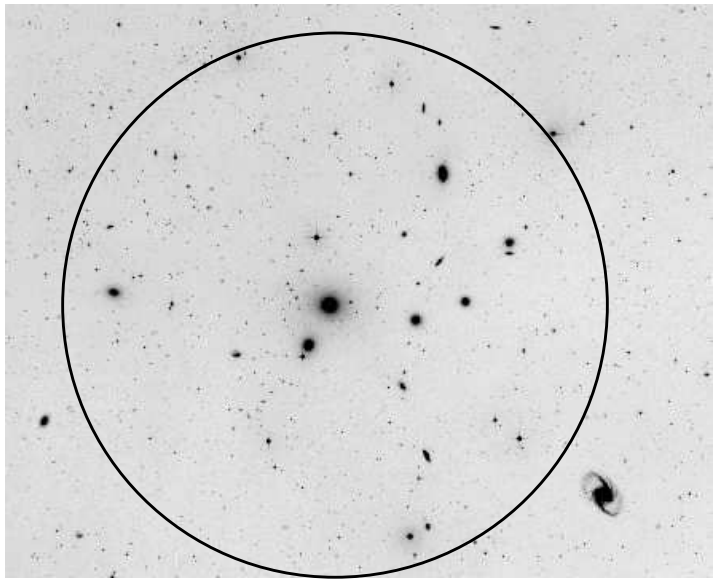}{90mm}{0}{60}{60}{-180}{-110}
\caption{The first 2dF Fornax field observed. We measured spectra of
2500 ``stars'' and ``galaxies'' with $B_J<19.7$ inside the 2 degree
diameter circle shown. The field is centred at R.A. $03^{\rm h}\;
38^{\rm m}\; 29^{\rm s}$, dec. $35^{\circ}\; 27'\; 01''$
(J2000).}\label{fig-field}
\end{figure}

In this paper we present some preliminary results from the FSS based
on the first season of observing in which we have almost completed our
first field, shown in Figure~\ref{fig-field}. Here we concentrate on
the LSB galaxies and in a companion paper (Drinkwater et al., this
volume) we present the detection of HSB galaxies from the survey.

\section{Observations}

The spectroscopic data were obtained with the 2dF in 1996 and 1997.
Full details of the observations and the other objects observed are
given in Drinkwater et al.\ (1999). In the limited time available
during these initial observations, we successfully observed 1041
(80\%) of the resolved objects to a limit of $B_J=19.7$ and a total of
1123 (45\%) of the stellar objects to the deeper limit of
$B_J=20.3$. Most of the observations were 2 hour exposures for the
galaxies and about 45 minutes for the stars.

\begin{figure}[!htb]
\plotone{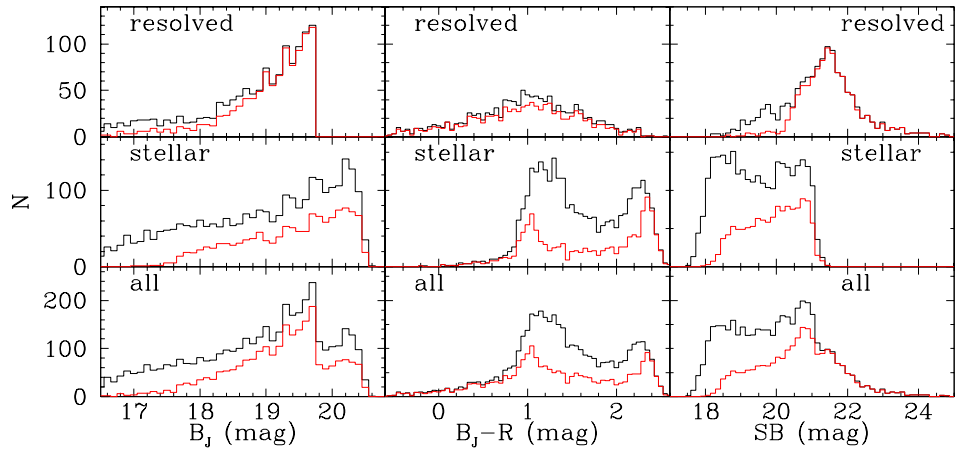}
\caption{The completness of our observations in the first 2dF field as
functions of magnitude, colour and central 
surface brightness for resolved
images, stellar images and all images combined. In each panel the
upper histogram indicates the total sample and the lower histogram
represents the numbers of objects observed. Note that the $R$
magnitudes are only reliable for stellar images, so the colours of the
resolved objects are only indicative. Similarly the surface brightness
estimates are not reliable for the stellar images, being determined by
seeing and saturation effects.}\label{fig-histo}
\end{figure}

The numbers of objects observed are shown in Figure~\ref{fig-histo} as
functions of their magnitude, colour and surface brightness for the
resolved and stellar image classes and for the whole sample
combined. The star--galaxy classification is taken from the APM
catalogue. The colours ($B_J-R_F$) were derived using the $R$
magnitudes from the APM catalogue data. These have not been
independently calibrated and are optimized for stellar images so the
galaxy colours are only indicative. However the stellar objects show
the normal colours with the classic bimodal distribution of blue halo
and red disk populations (Kron 1980).

The selection functions in Figure~\ref{fig-histo} show that, as
discussed above, we observed the stellar images to a fainter magnitude
limit than the resolved objects, with some bias not to observe the
brightest stars. The selection by colour was fairly uniform for
resolved objects, but the stellar objects were chosen with a bias to
extreme colours so as preferentially to include unusual objects.
There is a slight bias to faint surface brightness for all images
classes. We will complete this first field during out next observing
run to remove these biases.

\section{Analysis}

We reduced the data using standard IRAF routines for multi-fibre
spectroscopy, although an automated pipeline reduction package is now
also available at the AAT. We then cleaned the spectra by
interpolating across the strongest sky line residuals and correcting
for atmospheric absorption. We analysed all the spectra with the RVSAO
(Kurtz \& Mink 1998) cross-correlation package to identify the objects
and measure their radial velocities. Instead of the normal selection
of galaxy templates for the cross-correlations, we used a set of ten
stellar templates from the Jacoby, Hunter \& Christian (1984) library
plus one emission line galaxy template. These templates were capable
of identifying and measuring galaxy redshifts as well as a set of
galaxy templates, but had the advantage, when applied to Galactic
stars, of giving a good first estimate of the stellar type. The object
identifications were accepted if the Tonry \& Davis (1979) ``R''
coefficient was greater than 3, although all the identifications were
checked by visual inspection. If an object was not identified and
there were any signs of broad emission lines in the spectrum, it was
subsequently tested against a QSO template spectrum (Francis et al.\
1991). We used the same process to analyse all the spectra regardless
of the image morphology, to give object identifications based on the
spectra alone.

\begin{figure}[!thb]
\plotone{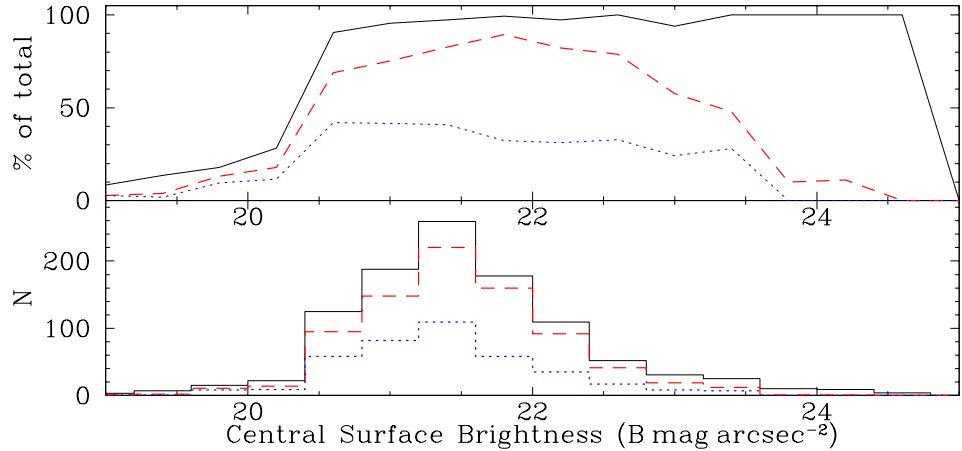}
\caption{The galaxies observed (solid), with measured redshifts
(dashed), and with strong emission lines (dotted) plotted as a
fraction of the total sample (upper panel) and as histograms of the
actual numbers (lower panel), both as a function of surface
brightness.}\label{fig-rate}
\end{figure}

Figure~\ref{fig-rate} indicates the success rate of the
identifications for galaxies in the sample (i.e.\ fraction of objects
with measured redshifts) as a function of surface brightness. This is
fairly constant at about 90\% except for the extremely LSB objects:
the success rate drops below 50\% fainter than 23 $B$\,mag
arcsec$^{-2}$.  We plan to extend our survey to fainter surface
brightness limits in future observations with one long (order 6 hour)
exposure in each field. This should permit us to get a reasonable
completeness to 23.5 $B$mag arcsec$^{-2}$. The Figure also shows that
about half the identified galaxies have strong emission lines. This
fraction increases to lower surface brightness, demonstrating the bias
against identifying absorption line spectra at low signal-to-noise 
ratios. 

\section{Results}

\begin{figure}[!thb]
\plotone{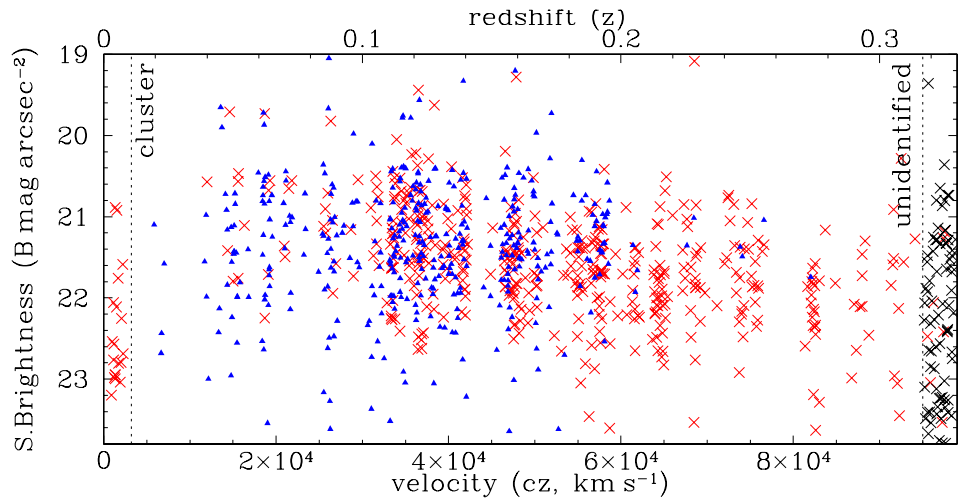}
\caption{The distribution in surface brightness and
redshift of all the galaxies observed so far. The triangles represent
objects measured with strong emission lines; all others are indicated
by crosses. Note that at redshifts above 55,000\,km\,s$^{-1}$ the H$\alpha$
emission line is shifted out of the 2dF spectra, so very few galaxies
are flagged for emission. Galaxies for which we could not measure a
redshift are plotted at the right of the figure.
}\label{fig-sb}
\end{figure}

In this section we summarise the initial results of the survey,
concentrating on the low surface brightness galaxies, but briefly
reviewing the other objects. Most of the galaxy results are summarised
in Figure~\ref{fig-sb} which shows the surface brightness plotted
against redshift for all the galaxies measured so far. Note that we
use the spectra to define which objects are galaxies: those with
redshifts greater than 700\,km\,s$^{-1}$, irrespective of morphology.

\subsection{Galactic stars} 

In addition to the galaxies shown in Figure~\ref{fig-sb}, we also
have a large sample of stars. These will be used for studies of
Galactic structure in later papers.

\subsection{Cluster galaxies}

Our galaxy observations were limited in both magnitude and surface
brightness, so only 20 galaxies classified as cluster members
by Ferguson (1989, FCC) were observed. Of these, 4 were actually found
to be background galaxies, the remaining 16 being confirmed as
members.

We detected 5 more members of the Fornax Cluster that were not listed
as cluster members in the FCC. Two of these new members were listed in
the FCC as background galaxies because they had relatively high
surface brightness (reported by Drinkwater \& Gregg, 1998). Examples
of these galaxies are shown in the top row of Figure~\ref{fig-images}.
All the cluster members observed, most of which have central surface
brightness $\mu_0>22.5B$\,mag\,arcsec$^{-2}$, were identified by
absorption line spectra, having no strong emission lines.  The
remaining 3 new members were not listed at all in the FCC, presumably
because they were too faint.  These new galaxies have red colours
($1.1<B_J-R<1.7$\,mag) and appear to be very small (scale lengths less
than 200\,pc), low luminosity dwarf ellipticals
($-11.9<M_B<-11.3$\,mag). We use 15.4\,Mpc as the Fornax cluster
distance and 30.9\,mag as the distance modulus (Bureau et al.\ 1996).
Some of the brighter new cluster dwarf galaxies are shown in the
middle row of Figure~\ref{fig-images}.

\begin{figure}[!thb]
\plotone{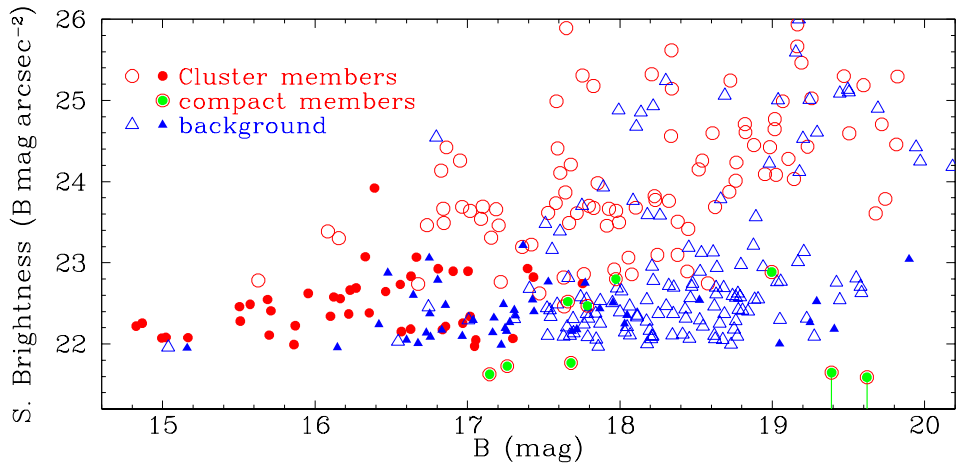}
\caption{Magnitude--surface brightness relation for galaxies in the
Fornax Cluster. The B band central surface brightness is plotted
against B magnitude. The membership classifications are taken from the
FCC (open symbols), except where they are now spectroscopically
confirmed (closed symbols). The two new compact galaxies indicated at
the lower right have surface brightnesses of 20.9.}\label{fig-dav}
\end{figure}

The detection of new cluster members like these weakens the
correlation between magnitude and surface brightness proposed by
Ferguson \& Sandage (1988) and refuted by Irwin et al.\ (1990). We
have replotted the data for this correlation in Figure~\ref{fig-dav}
with the new cluster members plotted. The scatter in the relation is
significantly increased, but the relation may still exist. 
We hope to resolve this with further observations.

\subsection{Field galaxies}

We denote as ``field galaxies'' all those galaxies beyond the Fornax
cluster (i.e.\ $cz>3500$\,km\,s$^{-1}$ but excluding those identified
as QSOs by having very broad emission lines. The field galaxies
include several high surface brightness compact emission line galaxies
which we have identified among the unresolved ``stellar''
objects. These are discussed in
more detail by Drinkwater et al.\ (this volume).

\begin{figure}[!thb]
\plotone{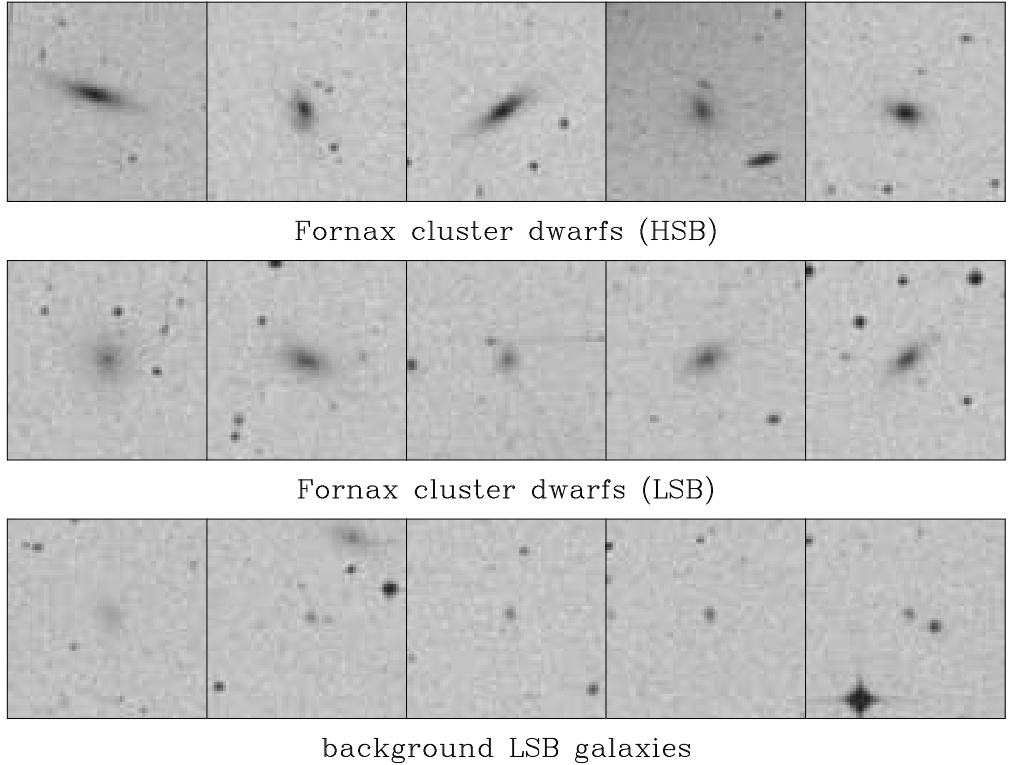}
\caption{$B_J$-band images of the new
galaxies. Each is 2$'$ across.
}\label{fig-images}
\end{figure}

It is apparent from Figure~\ref{fig-sb} that there is a tail of LSB
galaxies extending below the main galaxy population at all redshifts.
If we select the galaxies with central surface brightness
$\mu_0>22.5$\,B\,mag\,arcsec$^{-2}$ and redshift
$cz<50,000$\,km\,s$^{-1}$ (so H$\alpha$ is detectable), we note that
they {\em all} have emission line spectra and have bluer colours
($-0.4<B_J-R<0.7$\,mag) than the cluster LSBs. These field LSBs are
luminous ($-19.6<M_B<-17.0$\,mag for $H_0=75$\,km\,s$^{-1}$Mpc$^{-1}$)
and have relatively large scale lengths (2--5\,kpc) making them more
like local spirals (de Jong 1996) than dI galaxies.  A selection of
these LSB galaxies is shown in the bottom row of
Figure~\ref{fig-images}.

We also note from Figure~\ref{fig-sb} that there remains a large
number of LSB galaxies not yet identified (at the right edge of the
Figure). These galaxies are very interesting: if cluster members they
represent a significant new addition to the cluster luminosity
function. On the other hand if they are background field galaxies they
must include some very large galaxies.

\subsection{QSOs} 

We have so far identified 52 QSOs in the first field to a limit of
$B_J=20.3$, generally consistent with published QSO number counts. We
will eventually generate a large QSO sample with relatively free of
selection bias compared to other QSO surveys: for instance the current
sample covers a redshift range of $0.3<z<3.1$.

\section{Conclusions} 

In this paper we have presented results from the first 2500 spectra
measured for {\em The Fornax Spectroscopic Survey}.  We are continuing
this project with more observations scheduled in 1998 November when we
plan to extend the observed sample to fainter limits of surface
brightness. When complete the survey will comprise a unique sample of
the spectra of all 14,000 objects with $16.5<B_J<19.7$ in this
12\,deg$^2$ area of the Fornax Cluster. 
The ability to record spectra through 400 fibres simultaneously allows 
long exposures to be used while still sampling large numbers of galaxies. 
It is therefore possible to measure redshifts for the many low surface 
brightness galaxies observed in the direction of the Fornax Cluster, 
including both cluster members and background objects. Already new 
cluster members have been found, while other objects have been shown to 
be large field LSBGs in the background.

\end{document}